\documentclass{optica-article}

\journal{opticajournal}

\articletype{Letter}

\usepackage{gensymb}

\begin{document}

\title{Tunable and efficient ultraviolet generation in nanophotonic lithium niobate}

\author{Emily Hwang,\authormark{1,$\dagger$} Nathan Harper,\authormark{2,$\dagger$} Ryoto Sekine,\authormark{3} Luis Ledezma,\authormark{3,4} Alireza Marandi,\authormark{1,3} and Scott K. Cushing\authormark{2,*}}

\address{\authormark{1}Department of Applied Physics and Materials Science, California Institute of Technology, Pasadena, California, 91125, USA\\
\authormark{2}Division of Chemistry and Chemical Engineering, California Institute of Technology, Pasadena, California 91125, USA\\
\authormark{3}Department of Electrical Engineering, California Institute of Technology, Pasadena, California 91125, USA\\
\authormark{4} Jet Propulsion Laboratory, California Institute of Technology, Pasadena, CA 91109, USA\\
\authormark{$\dagger$} These authors contributed equally to this work.}

\email{\authormark{*} scushing@caltech.edu} 

\begin{abstract}
On-chip ultraviolet sources are of great interest for building compact and scalable atomic clocks, quantum computers, and spectrometers; however, few material platforms are suitable for integrated ultraviolet light generation. Of these materials, thin-film lithium niobate is the most competitive due to its ability to be quasi-phase-matched, optical confinement, and nonlinear properties. Here, we present efficient (197~$\pm$~5~\%/W/cm$^{2}$) second harmonic generation of UV-A light in a periodically poled lithium niobate nanophotonic waveguide. We achieve on-chip ultraviolet powers of 30~$\mu$W, demonstrating the potential for compact frequency-doubling of common near-IR laser diodes. By using a large cross section waveguide (600~nm film thickness), we achieve insensitivity to fabrication errors, and can attain first-order quasi-phase-matching with relatively long poling periods (>1.5~$\mu$m). The device also demonstrates linear wavelength tunability using temperature. By varying the poling period, we have achieved the shortest reported wavelength (355~nm) generated through frequency doubling in thin-film lithium niobate. Our results open up new avenues to realize ultraviolet on-chip sources and chip-scale photonics.
\end{abstract}

The field of integrated nonlinear optics has grown dramatically during the past decade due to the development and commercial availability of thin-film lithium niobate\cite{boes_lithium_2023,zhu_integrated_2021}. In passive nonlinear devices, thin-film lithium niobate (TFLN) excels in efficient frequency conversion and quantum state generation from the visible to the infrared (IR)\cite{nehra_few-cycle_2022,leidinger_comparative_2015}. The strong mode confinement of single-pass, low-loss\cite{shams-ansari_reduced_2022} nanophotonic waveguides and quasi-phase matched interactions utilizing lithium niobate’s largest second-order nonlinear optical tensor element have resulted in record-breaking efficiencies in applications such as second harmonic generation (SHG)\cite{park_high-efficiency_2022,jankowski_ultrabroadband_2020}, supercontinuum generation\cite{wang_ultrahigh-efficiency_2018}, difference frequency generation\cite{mishra_mid-infrared_2021}, parametric amplification\cite{ledezma_intense_2022}, and parametric downconversion\cite{zhao_high_2020}. Similarly, active devices such as modulators\cite{wang_integrated_2018}, electro-optic frequency combs\cite{zhang_broadband_2019}, and femtosecond pulse generators\cite{yu_integrated_2022} show impressive performance in compact form-factors due to lithium niobate’s large electro-optic tensor elements. However, there is still significant room for lithium niobate’s use in ultraviolet (UV) photonics\cite{soltani_alganaln_2016}, with applications such as UV-visible spectroscopy, optogenetics, high-resolution microscopy, UV security banknote features, laser cooling\cite{toyoda_laser_2001}, atomic clocks\cite{sugiyama_production_1997,ludlow_optical_2015}, and quantum computing\cite{saffman_quantum_2016}.

Although lithium niobate has been extensively studied in the IR, and comparatively less so in the visible, UV devices have remained rare to date. The few reported lithium niobate devices for UV generation have been limited to metasurfaces\cite{ma_nonlinear_2021}, nanoparticles\cite{timpu_lithium_2019}, and large micromachined or channel waveguides\cite{rutledge_broadband_2021,sugita_ultraviolet_2001}, and therefore do not take advantage of the high mode confinement and efficiency of lithium niobate in a nanophotonic platform. TFLN has yet to be well studied in the UV due to the ultra-short poling periods required to overcome the high dispersion in waveguides at short wavelengths, as well as material and scattering loss. Although lithium niobate’s band gap is near 315~nm (3.95~eV), an exponentially decaying Urbach absorption tail persists towards the visible due to defects in the crystal structure\cite{bhatt_urbach_2012}, and impurity ion (Cu$^{+}$, Fe$^{2+}$) resonances can cause additional loss\cite{schwesyg_light_2010,ciampolillo_quantification_2011}. Additionally, losses at the waveguide sidewalls increase at shorter wavelengths due to surface imperfection Rayleigh scattering, which scales as $\lambda^{-4}$\cite{maurer_glass_1973}. In spite of these difficulties, there is much to gain by extending the spectral coverage of TFLN frequency conversion to the UV. Notably, near-IR laser diodes, which can be frequency doubled, are considerably more accessible than UV laser diodes and gas lasers\cite{hasan_recent_2021}. Among other nanophotonic material platforms, only aluminum nitride (AlN) has been significantly investigated for waveguided second harmonic UV generation; however, AlN lacks ferroelectricity and is therefore incapable of periodic poling. The lateral polar structures used to achieve quasi-phase matching in AlN are highly scattering, resulting in much lower conversion efficiencies (<1\%)\cite{alden_quasi-phase-matched_2019} compared to lithium niobate devices. Other potential UV platforms (lithium tantalate\cite{meyn_fabrication_2001}, BBO\cite{devi_continuous-wave_2016}, LBGO\cite{umemura_temperature-dependent_2019}) have yet to be thoroughly explored in a thin film nanophotonic platform. Lithium niobate remains superior to these materials with its combination of low-loss waveguides, high second-order nonlinear response, ferroelectric poling for quasi-phase matching, and, for the case of thin-film lithium tantalate, commercial accessibility\cite{blumenthal_photonic_2020}.

\begin{figure}[htbp!]
\centering\includegraphics[width=12cm]{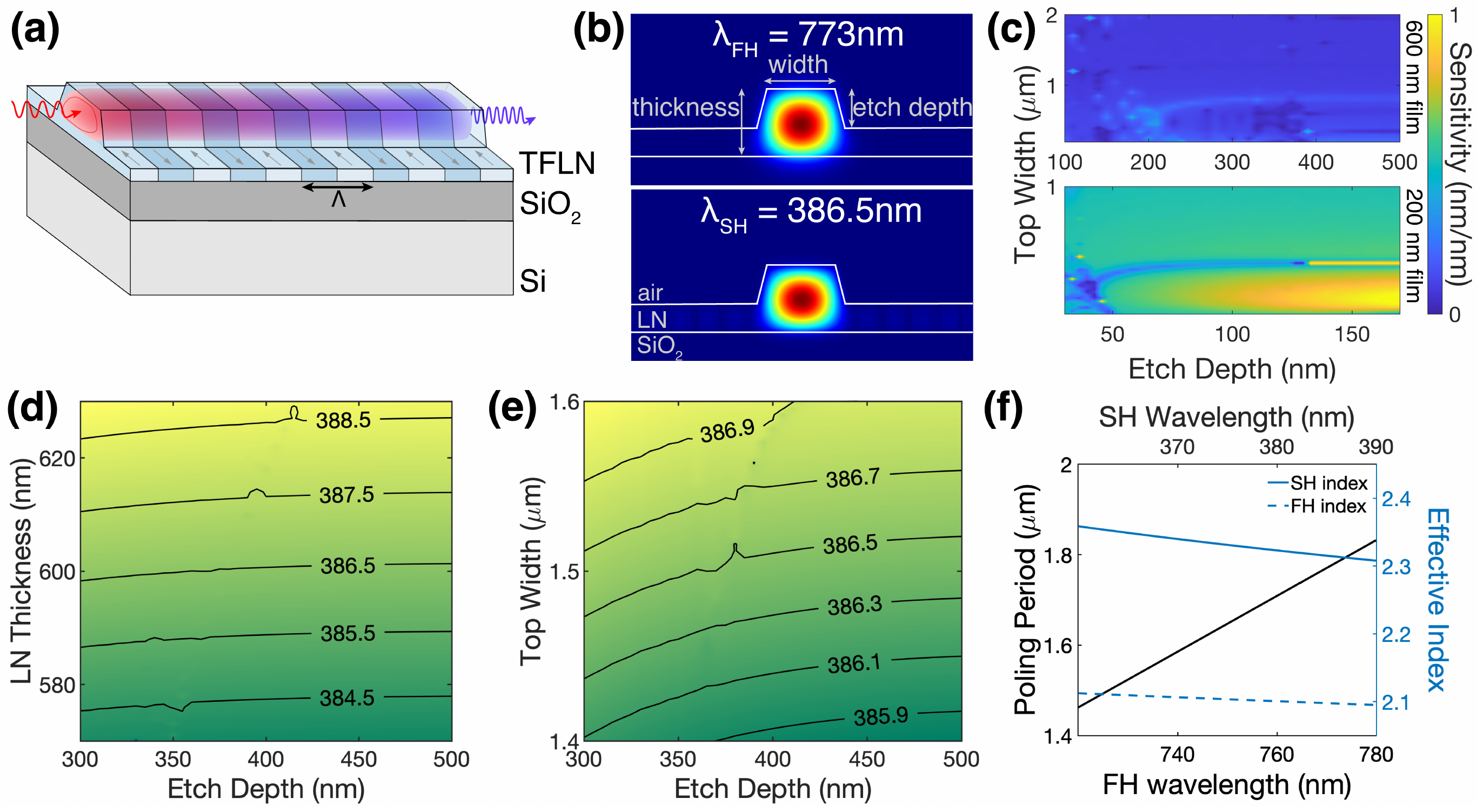}
\caption{\label{f:methods}(a) Schematic of the PPLN waveguide. (b) Mode profiles of the fundamental TE mode at the first and second harmonics. (c) Sensitivity of the SHG center wavelength to the thin film thickness as a function of the waveguide geometry for a 600~nm (top) and a 200~nm thin film thickness (bottom); local areas that minimize the sensitivity also decrease the phase matching bandwidth. Two dimensional sweeps of the (d) film thickness and etch depth and (e) top width and etch depth to vary the SHG center wavelength (contour lines). (f) Poling period and effective refractive indices of the first and second harmonic fundamental TE modes as a function of wavelength.}
\end{figure}

Here, we produce 30~$\mu$W of efficient (197~$\pm$~5~\%/W/cm$^{2}$) second harmonic generation of UV light (386.5~nm) with periodically poled lithium niobate (PPLN) rib waveguides (Fig.~\ref{f:methods}a). In addition to being among the first reported UV SHG devices in thin-film lithium niobate, the waveguides are designed to be robust against fabrication errors. The devices exhibits wavelength tunability through temperature and poling period, and is capable of UV SHG at the lowest wavelengths tested (710~nm frequency-doubled to 355~nm).

The optimal waveguide geometry was determined by minimizing the phase matching sensitivity to thickness variations in the LN thin film while maintaining the SHG bandwidth at the center wavelength of the laser used in this paper (773~nm). To date, variations in the thin film thickness of even 1~\r{A} are sufficient to disrupt phase matching in visible SHG, limiting the effective interaction length and chip-to-chip repeatability\cite{boes_lithium_2023,sayem_efficient_2021}. To quantify the phase matching sensitivity, the field profiles and effective indices of the guided modes were simulated (Lumerical MODE) using the bulk Sellmeier coefficients of lithium niobate\cite{gayer_temperature_2008} and SiO$_{2}$\cite{toyoda_temperature_1983} with the geometric parameters shown in Fig.~\ref{f:methods}b and a sidewall angle of 60\degree, which is consistent with the fabrication process. Only the fundamental quasi-transverse electric (TE) modes of the first harmonic (FH) and second harmonic (SH) (Fig.~\ref{f:methods}b) were considered since they access lithium niobate’s largest nonlinear tensor element (d$_{33}$~=~25~pm/V)\cite{shoji_absolute_1997}. The first-order wavelength sensitivity and SHG bandwidth were calculated using a numerical first derivative with respect to film thickness over a range of waveguide geometries for 200~nm and 600~nm thicknesses (Fig.~\ref{f:methods}c). For a waveguide top width of 1.5~$\mu$m and etch depth of 375~nm (Fig.~\ref{f:methods}d-e) with a 600~nm film, the wavelength sensitivity to film thickness is $\frac{d \lambda}{dT} =$ 0.08~nm/nm with a 3.4~pm SHG bandwidth, and the effective refractive indices ($n_{\text{eff,SH}}$~=~2.32, $n_{\text{eff,FH}}$~=~2.10) result in a quasi-phase matching poling period of $\Lambda=\frac{\lambda_{\text{SH}}}{(\Delta n_{\text{eff}})} =  1.8 \mu \text{m}$ (Fig.~\ref{f:methods}f). The larger cross section of the 600~nm film thickness allows for this relatively long poling period, in comparison to a 1.1~$\mu$m period for a 200~nm film with the same aspect ratio and wavelength. By using a waveguide with a larger cross section, we achieve insensitivity to the film thickness and fabrication errors without sacrificing the phase matching bandwidth.

The devices were fabricated from a 5\% MgO-doped X-cut thin-film lithium niobate on insulator wafer (NANOLN), which consists of 600~nm of lithium niobate bonded to 2~um of silicon dioxide on a 0.4~mm silicon substrate. Periodically poled waveguides with a 1.5~$\mu$m top width and 375~nm etch depth were fabricated following Ref.~\cite{ledezma_intense_2022}. Each waveguide had a 7~mm poled length, with poling periods ranging from 1.55 to 2~$\mu$m. The waveguide etch depth and sidewall angle were verified through atomic force microscopy, and the formation of poled domains was measured with second harmonic microscopy\cite{rusing_second_2019}.

\begin{figure}[htbp!]
\centering\includegraphics[width=12cm]{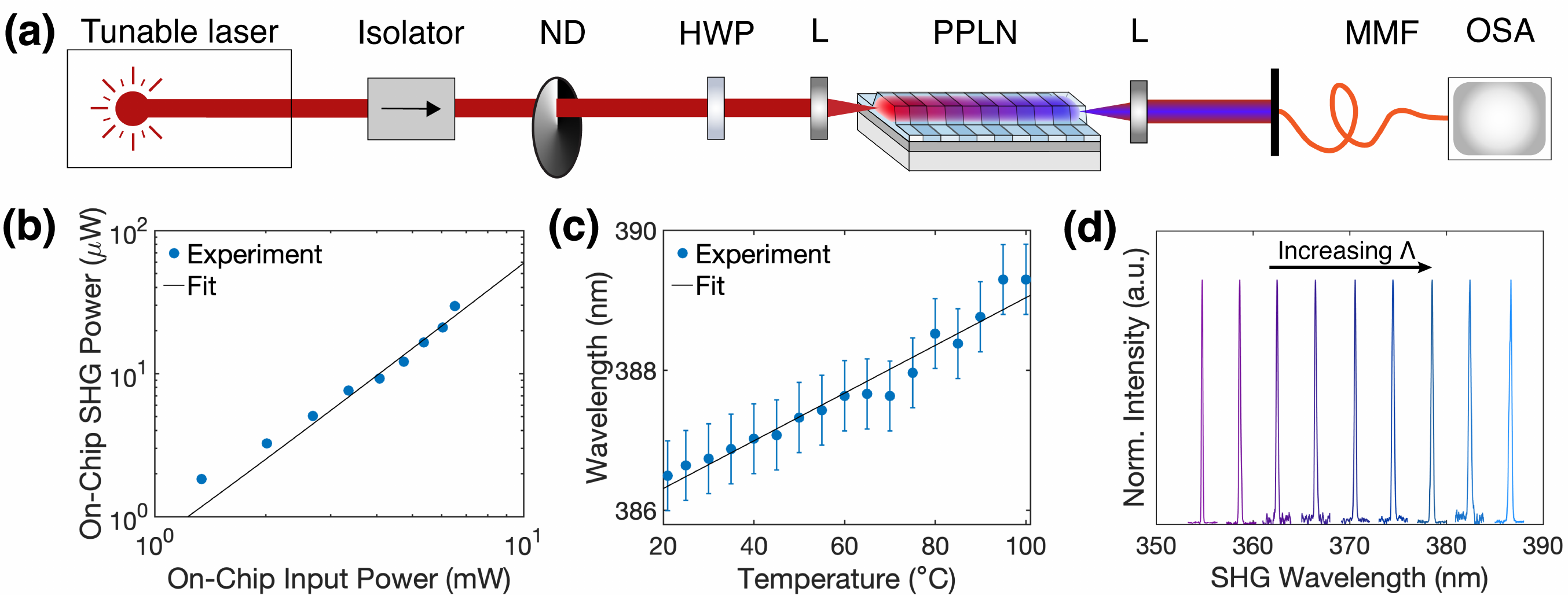}
\caption{\label{f:results} (a)~Optical setup for on-chip SHG characterization. ND, variable neutral density filter; HWP, half wave plate; L, aspheric lens; MMF, multi-mode fiber; OSA, optical spectrum analyzer. (b)~On-chip SHG output power measured as a function of the on-chip input power and experimental fit to calculate the efficiency. (c)~Temperature tuning of the SHG center wavelength, measured as the weighted average of the spectrum with corresponding standard deviations and experimental fit. (d)~Normalized spectra of the poling period sweep.}
\end{figure}

The SHG from the PPLN devices was characterized by the optical setup in Fig.~\ref{f:results}a. The output from a tunable continuous-wave (CW) single-frequency laser (Velocity TLD-6712, 765-781~nm) passed through an optical isolator and a variable neutral density filter to adjust the input power. An achromatic half-wave plate (Thorlabs AHWP10M-980) aligned the input polarization to the optical axis of the chip to maximize SHG power. The first harmonic was coupled to the waveguide using an AR-coated aspheric lens (focal length 1.5~mm, Thorlabs C140TMD-B). The waveguide output was collimated by another aspheric lens (focal length 1.5~mm, Thorlabs C140TMD-A) and collected by a high-OH multimode fiber (Thorlabs M122L01). The second harmonic was monitored using an optical spectrum analyzer (Yokogawa AQ6374) with a passband bandwidth of 5~nm around the second harmonic center wavelength to remove residual first harmonic.

The off-chip power of the first harmonic was varied from 6 to 30~mW at 773.1~nm, and the resulting on-chip SHG power is plotted in Fig.~\ref{f:results}b. We achieved a maximum on-chip UV power of 30~$\mu$W, and the slope of the least squares fit of the SHG output gives an efficiency of 197~$\pm$~5~\%/W/cm$^{2}$. The power law scaling suggests that photorefraction and pump depletion are not significant at the maximum power tested. The calculated efficiency and on-chip powers take the coupling losses into account, which were 22\% transmission per facet for the first harmonic and 10\% transmission per facet for the second harmonic. These values were determined by measuring the overall transmission at 780~nm and 405~nm, and assuming that the input and output coupling losses are equal. 

The temperature and poling period dependence of the phase matched wavelength is plotted in Fig.~\ref{f:results}c and \ref{f:results}d, respectively. To measure the temperature dependence, a nonlinear crystal oven (HC Photonics TC038-PC) heated the PPLN waveguides from 25 to 100~\degree C. The least squares slope (34~$\pm$~2~pm/\degree C) is in close agreement with the theoretically calculated value of 32~pm/\degree C. By varying the poling period, we also obtained SHG spectra extending down to 355~nm. A tunable CW Ti:Sapphire oscillator (Spectra-Physics Tsunami, 700-1100~nm) was used in a similar scheme to Fig.~\ref{f:results}a to measure the SHG from several PPLN waveguides with poling periods as short as 1.35~$\mu$m (Fig.~\ref{f:results}d). The fitted slope of the SHG center wavelength with the poling period (79.6~$\pm$~0.2~nm/$\mu$m) matches the theoretical slope of 81.1~nm/$\mu$m. This agreement in both the temperature and poling period demonstrate that the given Sellmeier coefficients\cite{gayer_temperature_2008} predict the temperature dependence and group velocity mismatch of lithium niobate in the UV relatively well, despite the limited refractive index data at these wavelengths. Unlike the initial measurements using the single-frequency laser, accurate efficiency data could not be extracted because the Ti:Sapphire oscillator linewidth (0.3~nm) is orders of magnitude larger than the phase matching bandwidth (3.4~pm). However, these spectra demonstrate that SHG is possible even closer to the 315~nm band gap of lithium niobate than what has previously been demonstrated, and that the thin film lithium niobate platform is able to phase match the full range of a Ti:Sapphire laser. 

\begin{figure}[htbp!]
\centering\includegraphics[width=10cm]{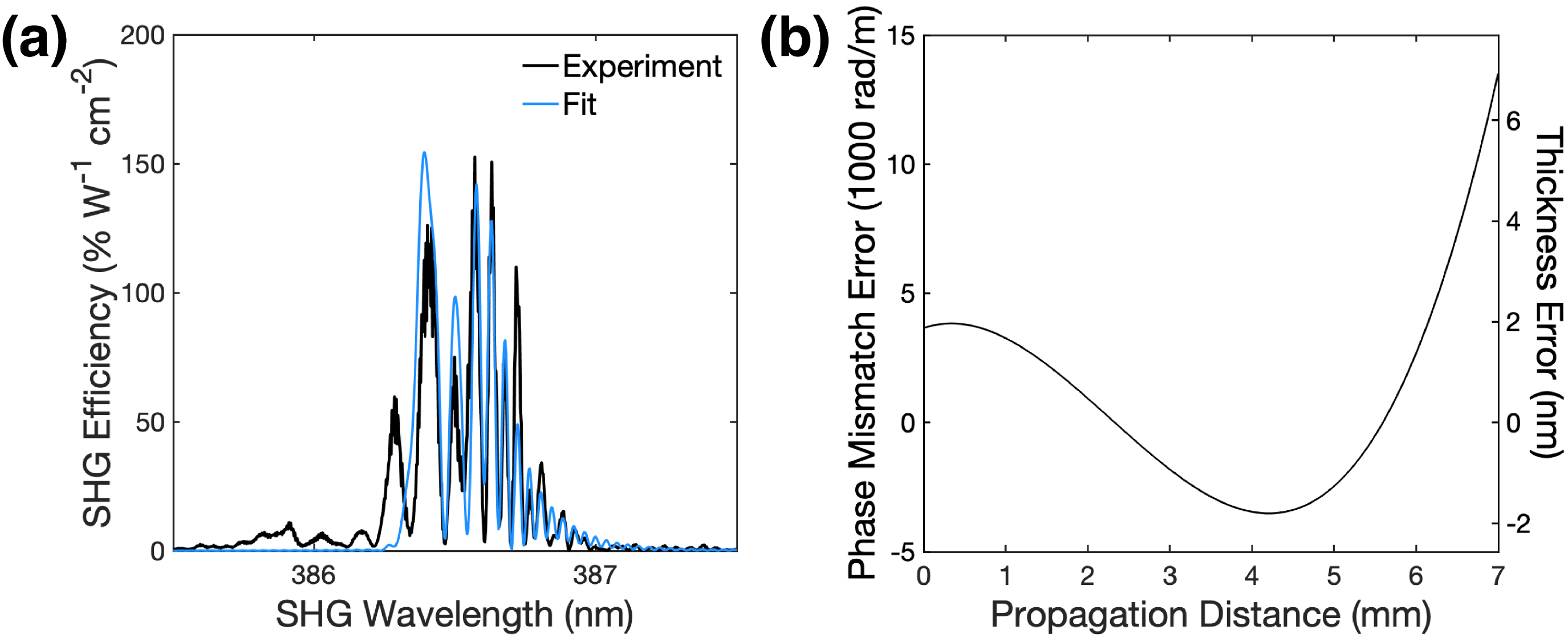}
\caption{\label{f:spectra} (a)~Experimental and index variation-fitted SHG spectra. (b)~Phase mismatch error represented by Eq.~\ref{eq:phase} and corresponding thickness error against propagation distance.}
\end{figure}

Although the SHG temperature and poling period tuning agree well with theory, the resulting 197~$\pm$~5~\%/W/cm$^{2}$ efficiency measured from the single-frequency laser at 773.1 nm is lower than the calculated theoretical efficiency\cite{jankowski_ultrabroadband_2020} of 18,100~\%/W/cm$^{2}$. The experimental SHG spectrum (Fig.~\ref{f:spectra}a) was also measured by sweeping the first harmonic wavelength from 771 to 775~nm with a constant input power of 30~mW. The spectrum exhibits multiple peaks over a 0.5~nm bandwidth, deviating from the theoretical sinc$^{2}$ line shape and 3.4~pm FWHM. Although the waveguides in this work support multiple modes at both the first and second harmonic, significant phase matching between higher-order modes is unlikely due to large momentum mismatches or poor modal overlap. Furthermore, the shape of the transfer function does not change when the second harmonic is collected with a single-mode fiber (Thorlabs P1-305A-FC-1), which is expected to occur if higher-order second harmonic modes were present. 

The discrepancies in the efficiency, bandwidth, and spectral shape are likely caused by index variations, potentially from thickness variations, thermal gradients, induced absorption, or photorefraction. Index variations preserve the area under the curve of the SHG transfer function, which allows us to determine the contribution of the index variations to the experimentally lower efficiency. An integral of the spectral efficiency yields an area of 31.8~\%/W/cm$^{2}\cdot$nm, and the transfer function with the same area and no index variations has a peak efficiency of 861~\%/W/cm$^{2}$. The poling quality of this device, which exhibited significant domain widening due to the short poling period, also contributes to the lowered efficiency. The duty cycle is estimated to be 90\% from second harmonic microscopy images, which lowers the theoretical efficiency to 1810~\%/W/cm$^{2}$. The remaining discrepancies can be explained by loss at the second harmonic and asymmetry in the input and output coupling efficiency. 

The magnitude of the index variations through the waveguide are estimated through simulation. Without any knowledge of the SHG phase relative to the first harmonic, the magnitude or position of the index variations cannot be directly calculated. However, the SHG spectrum can be approximated by the following expression for $\eta (\omega)$\cite{helmfrid_influence_1991}:
\begin{equation} \label{eq:eff}
    \eta(\omega) = \bigg\vert \int_{0}^{L} \text{exp}\left(i\int_{0}^{z} (\delta\beta(\xi) + \delta\beta(\omega)) \,d\xi\right) \,dz \bigg\vert ^{2}
\end{equation}
Where the phase mismatch error $\delta \beta(z)$ is expressed along the length of the waveguide as a cubic polynomial. The coefficients of $\delta \beta(z)$ are estimated by minimizing the squared residuals between the predicted and experimental spectra using a particle swarm optimization algorithm followed by a gradient descent\cite{kennedy_particle_1995,mezura-montes_constraint-handling_2011,pedersen_good_2010}, which yields the following expression for $\delta \beta(z)$ (Fig.~\ref{f:spectra}b):
\begin{equation} \label{eq:phase}
    \delta \beta(z) = 0.0636z^3  - 0.578z^2  + 0.544z + 3.66
\end{equation}
The blue trace in Fig.~\ref{f:spectra}a is the predicted spectrum using Eq.~\ref{eq:phase}. A thickness error of $\pm$5~nm throughout the waveguide explains the full spread in the experimental spectrum if the index variations are attributed only to thickness. A film thickness discrepancy of this magnitude is reasonable given the height measurements performed by NANOLN; however, this error is an upper estimate given that the index variations could be a combination of the thickness, photorefraction, UV-induced infrared absorption\cite{ali_observation_2011}, and temperature gradients. Although this thickness variation could be the cause of the lowered experimental efficiency, the same amount of thickness variation in a smaller film thickness could potentially disrupt phase matching entirely. The large cross sections of the waveguides makes the devices robust to the total film thickness, and therefore allows us to still achieve phase matching despite the lowered efficiency. Through this simple approximation, we believe that future devices can be optimized by estimating deviations from the ideal index, and that this method can serve as a means of monitoring photorefraction.

In conclusion, we have demonstrated fabrication-insensitive temperature-tunable ultraviolet light generation in an integrated thin-film lithium niobate waveguide. We have observed an experimental SHG efficiency of 197~$\pm$~5~\%/W/cm$^{2}$, with discrepancies from the theoretical efficiency that can be explained by the poling duty cycle and index variations in the waveguides. As of this publication, this is the shortest wavelength (355-386~nm) produced through second harmonic generation with periodically poled thin-film lithium niobate waveguides. Our work opens up opportunities to realize efficient frequency-doubled chip-scale ultraviolet laser diodes for UV integrated photonics, with applications spanning spectroscopy, atomic physics, and quantum science. 

\begin{backmatter}
\bmsection{Funding}
Department of Energy (DE-SC0020151); KNI-Wheatley Scholar in Nanoscience; National Science Foundation (DGE‐1745301); Rothenberg Innovation Initiative.

\bmsection{Acknowledgments}
The device fabrication was performed at the Kavli Nanoscience Institute at Caltech. The authors thank Robert Gray for his experimental support. E.H. was supported by the National Science Foundation Graduate Research Fellowship Program under Grant no. DGE‐1745301. Any opinion, findings, and conclusions or recommendations expressed in this material are those of the authors and do not necessarily reflect the views of the National Science Foundation. N.H. was supported by the Department of Defense (DoD) through the National Defense Science and Engineering Graduate (NDSEG) Fellowship Program. 

\bmsection{Disclosures}
The authors declare no conflicts of interest.

\bmsection{Data Availability Statement}
Data underlying the results presented in this paper are not publicly available at this time but may be obtained from the authors upon reasonable request.

\end{backmatter}


\bibliography{UVSHG}






\end{document}